\documentclass[twocolumn,preprintnumbers,floats,prd,amssymb,floatfix,nofootinbib,balancelastpage,superscriptaddress,amsmath]{revtex4-1}

\usepackage[utf8]{inputenc}
\usepackage{amssymb}
\usepackage{amsmath}
\usepackage{amsfonts}
\usepackage{graphicx}
\usepackage{color}
\usepackage{xspace}
\usepackage{comment}
\usepackage{hyperref}
\usepackage[normalem]{ulem}
\usepackage[section]{placeins}
\usepackage{afterpage}
\usepackage{slashed}
\usepackage{enumerate}
\usepackage{enumitem}
\usepackage{caption}
\usepackage{subfigure}
\usepackage{float}
\usepackage{ulem}
\usepackage{verbatim}
\usepackage{multirow,rotating}
\usepackage[dvipsnames]{xcolor}
\usepackage{braket}

\definecolor{dcolour}{rgb}{.5, .5, .5}
\def\gsim{\raise0.3ex\hbox{$\;>$\kern-0.75em\raise-1.1ex\hbox{$\sim\;$}}}
\def\lsim{\raise0.3ex\hbox{$\;<$\kern-0.75em\raise-1.1ex\hbox{$\sim\;$}}}
\def\gsim{\raise0.3ex\hbox{$\;>$\kern-0.75em\raise-1.1ex\hbox{$\sim\;$}}}
\def\lsim{\raise0.3ex\hbox{$\;<$\kern-0.75em\raise-1.1ex\hbox{$\sim\;$}}}

\newcommand{\ba}[1]{\begin{eqnarray} \label{(#1)}}
	\newcommand{\ea}{\end{eqnarray}}

\newcommand{\Eq}[1]{Eq.~\eqref{#1}}
\newcommand{\FIG}[1]{Fig.~\ref{#1}}
\newcommand{\TAB}[1]{Table~\ref{#1}}
\newcommand{\Threej}[6]{ \begin{pmatrix}
       #1 & #2 & #3 \\
       #4 & #5 & #6
\end{pmatrix}}

\newcommand{\Sixj}[6]{ \begin{Bmatrix}
       #1 & #2 & #3 \\
       #4 & #5 & #6
\end{Bmatrix}}

\newcommand{\Ninej}[9]{ \begin{Bmatrix}
       #1 & #2 & #3 \\
       #4 & #5 & #6 \\
       #7 & #8 & #9
\end{Bmatrix}}

\newcommand{\Comb}[2]{ \begin{pmatrix}
       #1 \\
       #2
\end{pmatrix}}

\begin{document}
\captionsetup[figure]{justification=raggedright,singlelinecheck=false}
\title{Solving two and three-body systems with deep neural networks}

\author{Ruitian Li}
\email[]{liruitian@mail.dlut.edu.cn}
\affiliation{Institute of Theoretical Physics, School of Physics, Dalian University of Technology, \\
		No.2 Linggong Road, Dalian, Liaoning, 116024, People’s Republic of China}
\author{Xuan Luo}
\affiliation{School of Physics and Optoelectronics Engineering, Anhui University, \\
		Hefei, Anhui 230601, People’s Republic of China}
\author{Hao Sun}
\email{haosun@dlut.edu.cn}
\affiliation{Institute of Theoretical Physics, School of Physics, Dalian University of Technology, \\
	No.2 Linggong Road, Dalian, Liaoning, 116024, People’s Republic of China}
\author{P. G. Ortega}
\email[]{pgortega@usal.es}
\affiliation{Departamento de F\'isica Fundamental, Universidad de Salamanca, E-37008 Salamanca, Spain}
\affiliation{Instituto Universitario de F\'isica
Fundamental y Matem\'aticas (IUFFyM), Universidad de Salamanca, E-37008 Salamanca, Spain}

\date{\today}

\begin{abstract}

We develop a new method for solving two- and three-body bound state problems using unsupervised machine learning techniques.
We use a deep neural network to calculate both simple and realistic potentials, obtaining the properties of the deuteron and triton bound states for the chiral effective field theory $NN$ potential. Our results provide significant accuracy with no prior assumptions about the behaviour of the wave function. This neural network technique, which extends from two-body to three-body, may provide insight into potential solutions to the nuclear and hadronic many-body problems.
\end{abstract}
\vskip10 mm
\maketitle
\flushbottom

\section{Introduction}
\label{I}

Over the years, the study of particle many-body systems has become one of the major research directions in physics. Particle many-body systems usually denote systems composed of multiple hadrons (e.g., protons and neutrons) that interact with each other through strong interactions to form complex collective behaviors. Solving the Schr\"odinger equation for these systems with high accuracy has remained one of the challenging problems in modern physics.

Many methods have been proposed to solve these systems. For instance, the Gaussian expansion method (GEM)~\cite{Kamimura:1988zz,Kameyama:1989zz,Hiyama:2003cu,Hiyama:2012gx,Hiyama:2012sma}, a method based on expanding the basis of the wave function in Gaussian series, has proven to be highly effective. However, when dealing with complex many-body systems, the need to construct and diagonalize large Hamiltonian matrices can significantly increase the computational complexity.
Other approximate methods, such as density functional theory and post-Hartree-Fock methods~\cite{Hohenberg:1964zz,Aspuru-Guzik:2018upc,Kirkpatrick:2021zhv}, have also been widely applied. However, these methods still exhibit limitations when extremely high precision is required.
Additionally, there exist other advanced techniques for many-body quantum systems, such as the quantum Monte Carlo method (QMC), including the variational Monte Carlo~\cite{McMillan:1965zz,Ceperley:1977zz,Ceperley:1980zz} and diffusion Monte Carlo (DMC)~\cite{von1992quantum,chin1990quadratic,needs2009continuum}. In particular, the variational Monte Carlo method deals with the ground state energy by using a variational approach, thus transforming the problem of finding the ground state of a system into a variational problem by means of a Monte Carlo integration. In all this methods, the accuracy of the final result depends on the initial parametrization of the trial wave function, that's it, on the initial ansatz of the wave function behaviour.

With the rise of artificial intelligence (AI) science, deep learning techniques, such as neural networks, have emerged as powerful complements to traditional computational approaches, offering new possibilities for solving a wide range of exponentially complex problems.
The most recent development in machine learning techniques is their application to solving specific physical problems in quantum mechanics~\cite{Dunjko:2018xgc,Mehta:2018dln,Carleo:2019ptp,Naito:2023piz,Wang:2024ynn,Wu:2025wvv,Keeble:2019bkv}.

The deep neural network (DNN) model of machine learning, which proceeds in networks by tuning the network parameters to minimize the loss function, has powerful optimization capabilities, thus providing an ideal tool for solving variational problems. Such techniques can potentially increase the accuracy of variational calculations, as they cover a much larger parametric space and allow to uniformly approximate any continuous function of real variables~\cite{Cybenko:1989iql,Hornik:1991sec}. For quantum many-body systems, the energy expectation $\langle H\rangle$ is treated as a loss function, whereas the output from machine learning is typically regarded as the wave function of the system under study, with no previous assumption on their structure.

In this work, we first analyze a DNN model to solve the two-body problem and later we extend it into the three body problem by using a multipole-expansion approach. In the DNN model, the coordinate variables are set as input, and the wave functions are the output.
The input layer and output layer are connected by hidden layers, each of which contains multiple nodes. Each node is connected to nodes in other layers through weighted inputs. We show that our method is successful to obtain the ground state of known examples of two and three body systems.

This paper is organized as follows: In Sec.~\ref{II}, we first describe neural networks and machine learning techniques and then discuss specific applications of machine learning techniques to two and three body systems. 
Then, in Sec.~\ref{III} are our numerical results and discussion.
Sec.~\ref{IV} is the summary of our work.

\section{THEORETICAL FRAMEWORK}
\label{II}
\subsection{Neural networks and machine learning techniques}

\begin{figure}[htbp]
	\centering
	\includegraphics[width=0.9\linewidth]{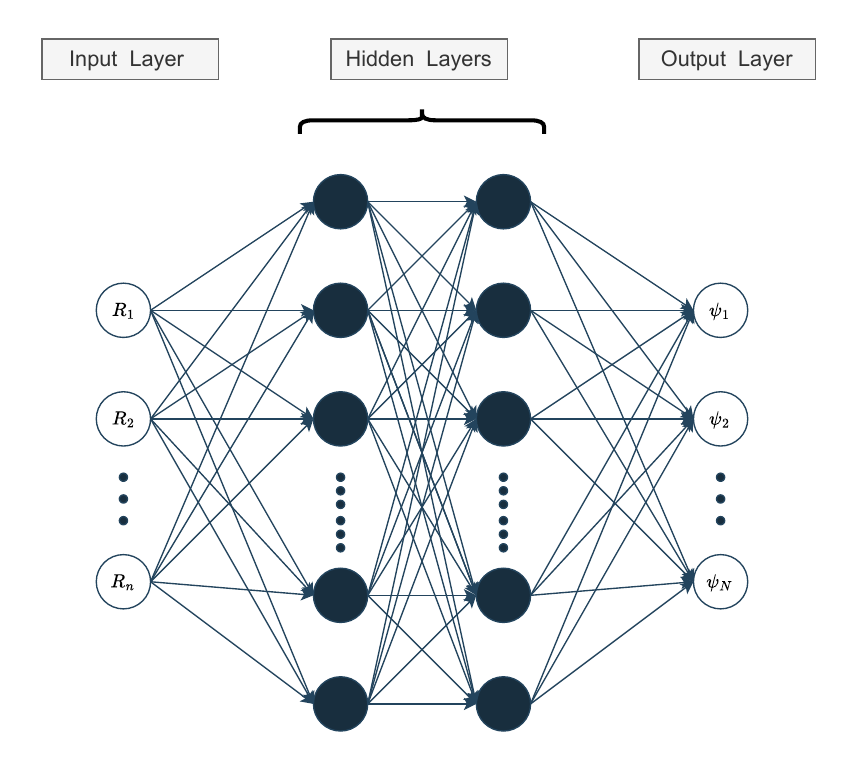}
	\caption{Diagram of the Deep Neural Network.}
	\label{f1}
\end{figure}

In this work we use deep neural networks (DNN) and machine learning techniques to obtain the wave functions and bound state energies of two and three body systems. Briefly, a deep neural network is a machine learning model inspired by the behaviour of biological neurons. They are powerful tools that allow to deal with complex tasks (image classification, speech recognition, search algorithms, etc.). In general, they consist on several layers of \emph{artificial neurons}, each one containing one or more binary inputs and one binary output~\cite{mcculloch1943logical}. They versatility comes when the neurons are organized in layers and connected among them as shown in \FIG{f1}. 

For our specific problem, the input layer is feed with the spatial coordinates $R=(R_1,\ R_2,\ \cdots, R_n)$ in which the space is discretized, where $n$ represents the number of coordinate degrees of freedom in the system,  and the output layer is the wave function $\psi(R)=(\psi_1(R), \ \psi_2(R), \ \cdots, \psi_N(R))$, where $N$ is the number of channels of the system.
The input and output layers are connected by a number of hidden layers, each with several nodes, as shown in \FIG{f1}. We divide the spatial coordinates $R_i$ equally into $M_i$ meshes, that is $R_i=(r_1,\ r_2,\ \cdots, r_{M_i})$.

The variational principle~\cite{10.1063/1.4869598} is used to find the ground state solutions of the many-body systems with Hamiltonian $H$:
\begin{equation}
	\langle H \rangle = \frac{\langle \Psi|H|\Psi \rangle}{\langle \Psi|\Psi \rangle} \geq E_0 ,
\end{equation}
which provides an approximate solution for the ground state wave function by minimizing the energy expectation value $\langle H\rangle$. Therefore, the expectation value of the system's ground state energy is a perfect candidate for the DNN's loss function.

In this paper, the parameters of the neural network needed to calculate the energy expectation $\langle H\rangle$ of the ground state are estimated through the machine learning software \textsc{Tensorflow}~\cite{tensorflow2015-whitepaper}, which is an optimized-performance state-of-the-art computing software based on Python.
For the loss function minimization, the gradient descent Adam optimizer~\cite{Kingma:2014vow} is employed. Adam, which stands for \emph{adaptive moment estimation}, is an optimizing algorithm that only requires first-order gradients of the parameters with little memory requirements. It computes individual adaptive learning rates for different parameters from estimates of the mean and variance of the data in an efficient way.

\subsection{Two-body system with DNN}\label{sec:twobody}

We will start considering a two-body system in the center of mass (CM) frame, so we only have one degree of freedom $\Vec{\boldsymbol{r}}$ which represents the separation between the two particles.
The corresponding two-body Schr{\"o}dinger equation is given by:
\begin{equation}\label{eqB1}
    H(r)\Psi(r,\theta,\phi) = E\Psi(r,\theta,\phi),
\end{equation}
where the Hamiltonian $H(r)$ consists of the kinetic term $T(r)$ and the potential term $V(r)$.
The total mass $M$ and the reduced mass $\mu$ of the system are
\begin{equation}
    M=m_{1}+m_{2}, \qquad \mu=\frac{m_{1}m_{2}}{M}.
\end{equation}
Assuming we are dealing with central potentials, we can use the spherical coordinate to express a separable wave function $\Psi(r,\theta,\phi)=\frac{u(r)}{r}Y_\ell(\theta,\phi)$.

Then the radial Schr{\"o}dinger equation is:
\begin{equation}\label{eqb5}
    \frac{-\hbar^2}{2\mu}\left[\frac{\partial^2}{\partial r^2}-\frac{\ell(\ell+1)}{r^2}\right]u(r) +(V(r)-E)u(r)=0 ,
\end{equation}
where $u(r)$ is the reduced radial wave function, $Y_\ell$ is the spherical harmonic function associated with the orbital angular momentum $\ell$ and $V(r)$ is the two-body potential.

As mentioned previously, in order to solve the two-body bound states using a DNN, the finite coordinate space $r$ in the input layer of the DNN network is divided into M meshes:
\begin{equation}
	r\approx \left(\begin{array}{c}
		r_1 \\
		r_2 \\
		\vdots  \\
		r_{M-1}\\
		r_{M}
	\end{array}\right),
\end{equation}
where each $r_i$ is uniformly distributed in space, with a spacing of $h$.

Hence, the centrifugal term and the interaction potential $V(r)$ for each partial wave are discretized and transformed into a diagonal matrix, as

\begin{equation}
	V_{ij} = V(r_i)\delta_{ij}
\end{equation}
with $\delta_{ij}$ the Kronecker delta.

For the second-order derivative of the wave function of \Eq{eqb5}, we approximate it by means of the central finite difference approximation,

\begin{align}\label{eq:discret}
\frac{\partial^2u(r_i)}{\partial r^2}  \approx \frac{u_{i+1}-2u_i+u_{i-1}}{h^2}
\end{align}
for each point $i$, which can be expressed in a matrix form as,

\begin{align}\label{eqb9}
\left(\frac{\partial^2u(r)}{\partial r^2} \right)_{ij} \approx \frac{1}{h^2}(-2\delta_{i,j}+1\delta_{i\pm 1,j})u_j,
\end{align}

The partial wave function, which is the output of the DNN final layer, is then predicted as the $M$-dimensional vector:
\begin{equation}
	u_i \equiv u(r_i) = \left(\begin{array}{c}
		u_1 \\
		u_2 \\
		\vdots  \\
		u_{M-1}\\
		u_{M}
	\end{array}\right).
\end{equation}

\subsection{Three-body system with DNN}\label{sec:threebody}

The rationale for solving the three-body system is similar to that for the two-body system, but new complications arise that need to be addressed.
The Schr{\"o}dinger equation for a three-body system consisting of three particles of mass $m_1$, $m_2$ and $m_3$ is:
\begin{equation}\label{eqc6}
    \left[(\sum_{i=1}^{3}\frac{-\hbar^2}{2m_i}\nabla^2_i)+\sum_{1=i<j}^{3}V(|{\bf r}_i-{\bf r}_j|)\right]\Psi=E\,\Psi ,
\end{equation}
where the index $i$ in the first term represents the kinetic energy of the particle $i$ and $V(|{\bf r}_i-{\bf r}_j|)$ is the potential between particle $i$ and particle $j$~\footnote{Particles will be indexed by $i$, $j$ and $k$, where $i\neq j$, $j \neq k$ and $i \neq k$.}.

\begin{figure}[t]
    \centering
    \includegraphics[width=0.9\linewidth]{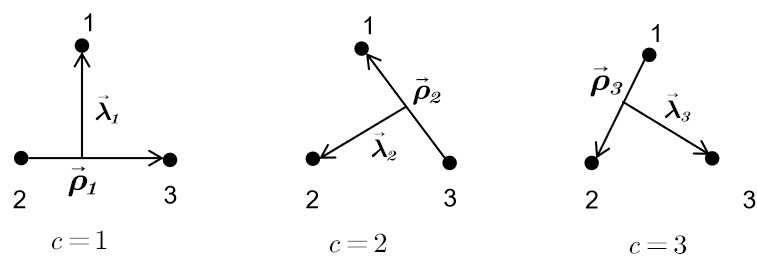}
    \caption{Jacobi coordinates in a three-body system.}
    \label{fc2}
\end{figure}

We will calculate the system in the center-of-mass frame, where ${\bf R}=M_T^{-1}\sum_{i=1}^3 m_i{\bf r}_i$ with $M_T=m_1+m_2+m_3$.
Then, in this frame the system can be described using two internal Jacobi coordinates: one as the relative distance between two particles $(ij)$, dubbed ${\boldsymbol \rho}$, and another as the relative distance between the particle $k$ and the center of mass of the $(ij)$ pair, called ${\boldsymbol\lambda}$. Depending on the chosen pair of particles we can build up to three different channels $(c=1-3)$ in Jacobi coordinates $({\boldsymbol{\rho}}_c,\ {\boldsymbol{\lambda}}_c)$, as shown in Fig.~\ref{fc2}, defined as:
\begin{equation}\label{eq:Jacobicoordinates}
	\begin{aligned}
		{\boldsymbol \rho}_c & = {\bf r}_i-{\bf r}_j\\
		{\boldsymbol \lambda}_c & ={\bf  r}_k -\frac{m_i{\bf r}_i+m_j{\bf  r}_j}{m_i+m_j},
	\end{aligned}
\end{equation}
where we take the permutations $(i,j,k)=(2,3,1)$ for $c=1$, $(3,1,2)$ for $c=2$ and $(1,2,3)$ for $c=3$.

As the three channels define the same system, we can transform one channel coordinate system into another as,

\begin{equation}
\begin{aligned}
    {\boldsymbol \rho}_i &= \alpha_{ij}\,{\boldsymbol \rho}_j + \beta_{ij}\,{\boldsymbol \lambda}_j,\\
    {\boldsymbol \lambda}_i &= \gamma_{ij}\,{\boldsymbol \rho}_j + \delta_{ij}\,{\boldsymbol \lambda}_j,
    \end{aligned}
\end{equation}
with

\begin{equation}
    \begin{aligned}
        \alpha_{ij} &= \delta_{ji} = -\frac{m_i}{m_i+m_k},\\
        \beta_{ij} &= \epsilon_{ijk},\\
        \gamma_{ij} &= -\epsilon_{ijk}\frac{m_k(m_i+m_j+m_k)}{(m_i+m_k)(m_j+m_k)},
    \end{aligned}
\end{equation}
for $i\ne j$ and $\alpha_{ii}=\delta_{ii}=1$, $\beta_{ii}=\gamma_{ii}=0$, where $\epsilon_{ijk}$ is the usual Levi-Civita symbol.

Thus, in the center-of-mass frame, the Schr\"odinger equation of Eq.~\ref{eqc6} can be written as:
\begin{equation}\label{eqc12}
	\left[-\frac{\hbar^2}{2\mu_{\rho}}\nabla_\rho^2 - \frac{\hbar^2}{2\mu_{\lambda}}\nabla_\lambda^2  + V_T\right]\Psi=E\Psi,
\end{equation}
where the corresponding reduced masses are:
\begin{equation}
		\mu_{\rho}=\frac{m_im_j}{m_i+m_j}\quad \text{and}  \quad \mu_{\lambda}=\frac{m_k(m_i+m_j)}{m_i+m_j+m_k},
\end{equation}
and the local operator potential $V_T$ is given by:
\begin{equation}
	V_T =\sum_{1=i<j}^{3}V(|{\bf r}_i-{\bf r}_j|) = V({\boldsymbol\rho}_1) + V({\boldsymbol\rho}_2) + V({\boldsymbol\rho}_3).
\end{equation}

An usual approach to solve the three-body Schr\"odinger equation is to build a total wave function which is a combination of the three rearrangement channels $c=1-3$,

\begin{equation}\label{eqc7}
    \Psi_{J}=\Psi^{(1)}_{J}({\boldsymbol \rho}_1,{\boldsymbol \lambda}_1)+\Psi^{(2)}_{J}({\boldsymbol \rho}_2,{\boldsymbol \lambda}_2)+\Psi^{(3)}_{J}({\boldsymbol \rho}_3,{\boldsymbol \lambda}_3).
\end{equation}
However, this approach forces to deal with the three channel coordinate systems at the same time, which is not convenient for the machine learning approach considered in this work where we want to predict the total wave function in an unique $\rho$-$\lambda$ grid. Alternatively, we will select one rearrangement channel, e.g. $c=1$, and make a multipole expansion of the $V({\boldsymbol\rho}_c)$ potentials and $\psi_{J}^{(c)}({\boldsymbol\rho}_c,{\boldsymbol\lambda}_c)$ wave functions (with $c=2,3$) in terms of ${\boldsymbol \rho}_1$ and ${\boldsymbol \lambda}_1$~\footnote{From now on, we will just denote the variables in channel $c=1$ as ${\boldsymbol \rho}$ and ${\boldsymbol\lambda}$.}.

Hence, as detailed in Appendix~\ref{sec:multipoleNN}, the general potential $V({\boldsymbol\rho}_c)$ can be expanded in terms of $\rho$ and $\lambda$ variables as,

\begin{align}
 V_k(\rho_c) &= \sum_{l_1,l_2}\,V_{(l_1,l_2;k)}(\alpha_{c1}\rho,\beta_{c1}\lambda)\,[Y_{l_1}(\hat \rho)\otimes Y_{l_2}(\hat \lambda)]_k
\end{align}
where $k=0,1,2$ for central, spin-orbit and tensor potentials, respectively.

The total three body total wave function can be factorized as

\begin{align}
    \Psi_{J}^{(\alpha)} = \sum_c \Psi^{(c,\alpha)}_{L}\xi_{STC}^{(\alpha)}
\end{align}
where $\xi^{(\alpha)}_{STC}$ is the spin-isospin-color wave function and $\alpha$ encodes all the quantum numbers necessary to define a partial wave coupled to a global $\left(I\right)J^P$~\footnote{Henceforth we will drop the $\alpha$ index for simplicity.}.
For $c=\{2,3\}$, the orbital partial wave function $\Psi^{(c)}_{L}$ are expanded as detailed in Appendix~\ref{eq:multiplePsi},

\begin{align}
  \Psi^{(c)}_{L}({\boldsymbol \rho},{\boldsymbol\lambda}) &= \sum_{l_i} \Phi(l_i)\, \overline{\psi}(l_x,l_y;\rho,\lambda)\,\left[Y_{l_a}(\hat \rho)Y_{l_b}(\hat \lambda)\right]_{L}
\end{align}
where $l_i$ represents all internal orbital momenta different from $L$ and $\Phi$ and $\overline{\psi}$ are defined in  Eqs.~\ref{eq:PhiME} and~\ref{eq:PsiME}, respectively.
For $c=1$, this reduces to
\begin{align}
    \Psi^{(1)}_{L}({\boldsymbol \rho},{\boldsymbol\lambda}) &= \psi^{(1)}(\rho,\lambda)\,[Y_{l_\rho}(\hat \rho)\otimes Y_{l_\lambda}(\hat\lambda)]_{L},
\end{align}
being $l_{\rho(\lambda)}$ the orbital angular momentum of the $\rho$($\lambda$)-mode, coupled to the total orbital momentum $L$.

For convenience we will define the reduced radial wave function,
\begin{align}
    \psi^{(c)}(\rho,\lambda) &=\frac{u^{(c)}(\rho,\lambda)}{\rho\,\lambda}.
\end{align}

The spatial matrix elements of the interactions for different partial waves are, then, calculated as
\begin{equation}
		\langle V_k(\rho_c)\rangle_{\alpha'\alpha} = \langle\Psi^{(\alpha')}_{L'M'}({\boldsymbol\rho},{\boldsymbol\lambda}) | V_k( \rho_c) |\Psi^{(\alpha)}_{LM}({\boldsymbol\rho},{\boldsymbol\lambda})\rangle.
\end{equation}

These matrix elements are completed with the spin-isospin recoupling coefficients, which can be conveniently represented by the Wigner 6-j symbol:
\begin{equation}
    \begin{aligned}
        {\cal S} &\equiv \langle s_1(s_2s_3)s_{23};S|(s_1s_2)s_{12}s_3;S\rangle \\
        =& (-1)^{s_1+s_2+s_3+S}\sqrt{(2s_{12}+1)(2s_{23}+1)}\begin{Bmatrix} s_1& s_2& s_{12}\\s_3 &S&s_{23}\end{Bmatrix}
    \end{aligned},
\end{equation}
and similarly for the isospin recoupling $\cal{I}$, for which it is sufficient to replace spin with isospin, e.g., $S\to T$ and $s_i\to t_i$.

Regarding the kinetic term, the $\nabla_\rho^2$ and $\nabla_\lambda^2$ terms in Eq.~\eqref{eqc12} can be directly expressed as:
\begin{equation}
	\nabla_\rho^2 \psi= \frac{1}{\rho\,\lambda}\left(\frac{\partial^2}{\partial \rho^2} - \frac{l_\rho(l_\rho+1)}{\rho^2}\right)u,
\end{equation}
and similarly for the $\lambda$ variable, which are discretized following the same procedure as in Eq.~\eqref{eqb9}.

As we focus on a three-body bound state, we need to divide the finite space represented by two coordinates $R=(\rho,\lambda)$ into $M_{\rho}$ and $M_{\lambda}$ meshes. So the total number of meshes in the DNN input layer is $M_{\rho} \times M_{\lambda}$. The predicted discretized partial wave function $u^{(n)}$ can be, then, linearized as
\begin{equation}
	u(\rho,\lambda) \approx \left(\begin{array}{c}
		u_{11} \\
		u_{12} \\
		\vdots  \\
		u_{1{M_{\lambda}}}\\
		u_{21}\\
		\vdots\\
		u_{{M_{\rho}(M_{\lambda}-1)}}\\
		u_{{M_{\rho}M_{\lambda}}}
	\end{array}\right),
\end{equation}
in an equivalent formalism as for the two body case detailed in Sec.~\ref{sec:twobody}.

\section{RESULTS AND DISCUSSIONS}
\label{III}
\subsection{Two-body systems}

\subsubsection{Study of the two-body isotropic harmonic oscillator}

For an initial test the two-body DNN code we will solve the 3D isotropic harmonic oscillator, whose potential is,
\begin{align}\label{eq:pot2BHO}
 V = \frac{1}{2}\mu\omega^2 r^2
\end{align}

The solution of this potential can be found in any Quantum Mechanics textbook~\cite{Sakurai:1167961}, having the general solution

\begin{align}\label{eq:2bHOsol}
\Psi({\bf r}) = {\cal N} r^\ell e^{-\frac{1}{2}\mu\omega r^2} L_k^{(\ell+1/2)}(\mu\omega r^2) Y_{\ell m}(\hat{r})
\end{align}
with $\cal{N}$ the normalization of the wave function and $L_k^{(\ell+1/2)}$ the associated Laguerre polynomials. The energy of the states is given by,

\begin{align}
 E=\hbar\omega\left(n+\frac{3}{2}\right)
\end{align}
with $n=2q+\ell$ for $q=0,1,2,\ldots$ and $\ell=0,1,2,\ldots$.

For the numerical comparison, we will take $\ell=0$, $\mu=500$ MeV and $\hbar\omega=1$ MeV, which gives a ground state of mass $E=1.5$ MeV and an analytical wave function given by,
\begin{align}\label{eq:analytic2bHO}
 \Psi(r) &= \left(\frac{2\beta}{\pi}\right)^{3/4} e^{-\beta r^2}
\end{align}
with $\beta = 250$ MeV$^2$.

\begin{figure}[t]
 \includegraphics[width=.5\textwidth]{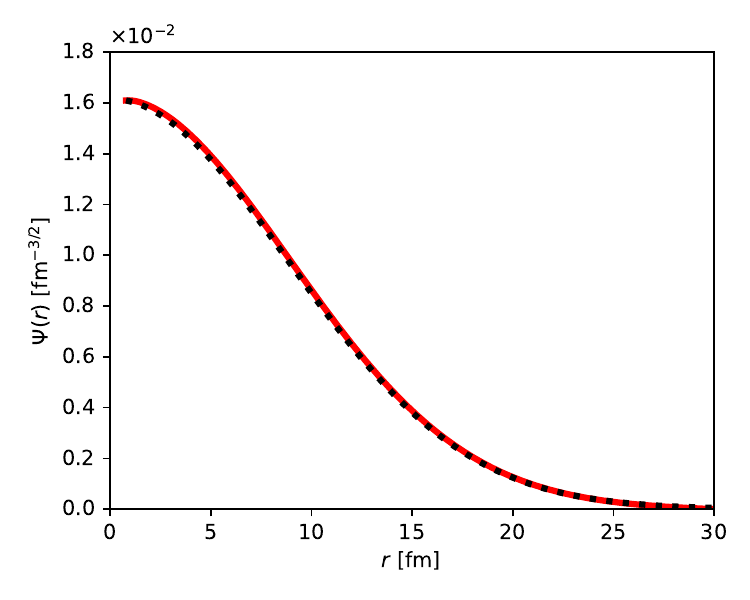}
 \caption{\label{fig:2BHOwf} DNN-predicted wave function (red line) and exact wave function of Eq.~\eqref{eq:analytic2bHO} (black dotted line) for the two-body isotropic harmonic oscillator. The difference between the two curves is found to be smaller than $2\%$. }
\end{figure}

Following the procedure described in Sec.~\ref{sec:twobody}, we build a DNN with two-hidden layers of $16$ nodes each. The radial coordinate is discritized using a $M=200$ mesh between $0.01$ and $30$ fm. The softplus function
 \begin{equation}
 	\text{softplus}(x) = \beta^{-1}\text{log}(e^{\beta x} + 1)
 \end{equation}
is used as the activation function of the neural network, with $\beta=1$ by default. This activation function is a smooth approximation to the popular ReLU function (ReLU$(x)=\max(0,x)$)~\cite{householder1941theory}, and is adequate when the output of the neural network is expected to be always positive in the considered space mesh. The output of the DNN gives a wave function which deviates less than $2\%$ of the exact wave function, as shown in \FIG{fig:2BHOwf}, predicting an energy of $1.5018$ MeV for the ground state. As expected, the virial theorem is also fulfilled, giving an average kinetic energy of $\langle T\rangle=0.7510$ MeV and an average potential energy $\langle V\rangle=0.7508$ MeV.

\subsubsection{Study of the deuteron}\label{sec:deuteron}

We now analyze the deuteron case, which involves a $^3S_1$-$^3D_1$ coupled channels calculation and more complex potentials. The deuteron consists of a proton and a neutron and is the simplest two-nucleon bound state system. Until now, the deuteron has been intensively studied both experimentally and theoretically, and the relevant experimental data are very precise. These well-established dates provide a reliable benchmark for studying nuclear systems using machine learning methods. Therefore, we calculated deuteron to examine the efficacy of our approach.

For that, we use the chiral effective field theory ($\chi$EFT) approach at N$^3$LO~\cite{Entem:2003ft,Entem:2002sf}  derived from the principles of quantum chromodynamics (QCD) to represent the nuclear force between two nucleons using chiral perturbation theory.
The degrees of freedom relevant for Chiral EFT are pions and nucleons. The nucleon-nucleon (NN) interaction is described by the short-range contact terms and long-range pion exchanges.
Then, the potential arise in terms of contributions to the momentum-space NN amplitudes in the CM system, which can be generally decomposed as
\begin{equation}\label{eq:NNpot}
	\begin{aligned}
		V\left(\vec{p}\,', \vec{p}\right)= & V_C+\boldsymbol{\tau}_1 \cdot \boldsymbol{\tau}_2 W_C \\
		& +\left[V_S+\boldsymbol{\tau}_1 \cdot \boldsymbol{\tau}_2 W_S\right] \vec{\sigma}_1 \cdot \vec{\sigma}_2 \\
		& +\left[V_{L S}+\boldsymbol{\tau}_1 \cdot \boldsymbol{\tau}_2 W_{L S}\right][-i \vec{S} \cdot(\vec{q} \times \vec{k})] \\
		& +\left[V_T+\boldsymbol{\tau}_1 \cdot \boldsymbol{\tau}_2 W_T\right] \vec{\sigma}_1 \cdot \vec{q} \vec{\sigma}_2 \cdot \vec{q},
	\end{aligned}
\end{equation}
where $\vec p\,'$ and $\vec p$ represent the final and initial nucleon momentum, respectively, and operators $\vec \sigma_{1,2}$ and $\boldsymbol{\tau}_{1,2}$ correspond to the spin and isospin of nucleons 1 and 2, respectively.
The average momentum is denoted by $\vec k = (\vec p\,' + \vec p)/2$, and the total spin is given by $\vec S = (\vec\sigma_1 + \vec\sigma_2)/2$. The subscripts $C$, $S$, and $T$ refer to the central, spin-spin, and tensor components, respectively, with $V$ denotes the isoscalar potential and $W$ the isovector potential.
We will work with the corresponding position-space potentials, obtained by applying the Fourier transform to the momentum-space function as described in Ref.~\cite{Saha:2022oep}.

For $J=S=1$ channels, the tensor force contributes to the NN potential, and the \Eq{eqb5} becomes two coupled equations for the $^3S_1$ and $^3D_1$ waves.
Then, the Schr{\"o}dinger equation can be written as:
\begin{equation}
	\begin{aligned}
			&\left[ -\frac{\hbar^2}{2\mu}\frac{\partial^2}{\partial r^2}\begin{pmatrix} 1&0\\0&1\end{pmatrix}  + \begin{pmatrix} V_{SS}& V_{SD} \\ V_{DS} & (V_{DD}+ \frac{\hbar^2}{2\mu}\frac{6}{r^2}) \end{pmatrix}  \right] \begin{pmatrix}u_S(r)\\  u_D(r)\end{pmatrix}\\
		&= E \begin{pmatrix}u_S(r)\\  u_D(r)\end{pmatrix}
	\end{aligned}\label{eqb7},
\end{equation}
where $u_S(r)$ and $u_D(r)$ are the reduced radial wavefunctions for the $S$ and $D$-waves, respectively, and $V_{SS}$, $V_{SD}$, $V_{DS}$ and $V_{DD}$ are the NN potentials of the two-particle partial-wave decomposition.

\begin{figure}[htbp]
	\centering
	\includegraphics[width=0.45\textwidth]{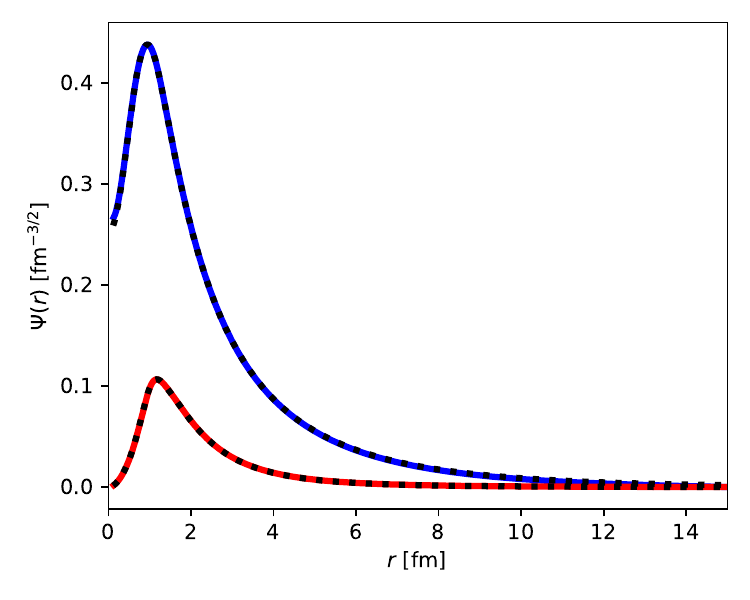}
	\caption{Deuteron radial wave functions $\Psi_S(r)$ (blue line) and $\Psi_D(r)$ (red line) for $n_{h1,2}=16$ nodes and $M=1000$ points. For comparison, the dotted black lines show the wave functions obtained from the Gaussian Expansion Method (GEM).}
	\label{fig3}
\end{figure}

It is well known that the ground state of deuteron is a proton and neutron state with quantum numbers of $S=J=1,T=0$, and its composition is a mixture of around $96\%$ $^3S_1$ state and $4\%$ $^3D_1$ state~\cite{Bertulani:2007bfy,Dumbrajs:1983jd}.
We calculated these two partial wave functions in an uniform mesh between $0.01$ fm and $15$ fm with $M=1000$ points. The DNN used consists of two hidden layers, each containing $16$ nodes with the softplus activation function.
We show in \FIG{fig3} the deuteron wave functions, where $\Psi_S(r)=u_S(r)/r$ and $\Psi_D(r)=u_D(r)/r$ denote the wave functions for the $^3S_1$ and $^3D_1$ states, respectively. For comparison, we also show the wave functions obtained from the Gaussian Expansion Method (GEM)~\cite{Hiyama:2003cu} with $n=40$ gaussians, $r_{\rm min}=0.01$ fm and $r_{\rm max}=15$ fm.

The neural network predicts a state with a binding energy of $2.2241$ MeV, in good agreement with the experimental value. \TAB{deuteron} lists the predicted deuteron properties based on the wave functions obtained from DNN and compared to empirical information from Refs.~\cite{VANDERLEUN1982261,PhysRevC.51.1127} and the calculations of Ref.~\cite{Saha:2022oep}. Furthermore, we also provide the probability of the deuteron $D$-wave state, of $4.10$ \%. Although this quantity is not directly observable, it is of significant theoretical importance.
The results are consistent with the well-established properties of the deuteron (see Table~\ref{deuteron}) where the deuteron ground state consists predominantly of the $^3S_1$ state, with the $^3D_1$ state accounting for only a small fraction.

\begin{table}
	\captionsetup{justification=raggedright, singlelinecheck=false}
	\caption{ The DNN prediction of deuteron properties (binding energy, matter radius and D-wave probability) compared to N$^3$LO $\chi$EFT and experimental data.}
 	\begin{tabular}{c|cccc}
 		\hline\hline
 		   &  DNN  & $\chi$EFT~\cite{Saha:2022oep} & Exp.~\cite{VANDERLEUN1982261,PhysRevC.51.1127}\\
 		  \hline
 		 $E_b$ [MeV]& $2.2241$ &  $2.22458$ & $2.224575(9)$ \\
 		 $r_d$ [fm] & $1.867$ & $-$ & $1.971(6)$\\
 		 $\cal{P}_D $ [\%]& $4.10$ & $4.03$ & $-$ \\
 		\hline\hline
    \end{tabular}
\label{deuteron}
\end{table}

\begin{table}[htbp]
	\centering
	\caption{Sensitivity of the deuteron results with the number of points in mesh, using $n_h=16$ nodes in each hidden layer.}
	\begin{tabular}{cccccc}
		\hline\hline
		$M$& $E_b$ [MeV] & $\langle T\rangle$ [MeV] & $\langle V \rangle$ [MeV] & ${\cal P}_S$ [\%] & ${\cal P}_D$ [\%]\\
		\hline
		$100$& $2.5340$ & $11.9729$ & $-14.5069$ & $95.80$ & $4.20$\\
		$200$& $2.3705$ & $11.8113$ & $-14.1818$ & $95.86$ & $4.14$\\
		$500$& $2.2624$ & $11.7334$ & $-13.9962$ & $95.90$ & $4.10$\\
	   $1000$& $2.2241$ & $11.7242$ & $-13.9479$ & $95.90$ & $4.10$\\
		\hline
		\hline
	\end{tabular}%
	\label{tab1}
\end{table}%

To analyze the convergence of the DNN results, we conducted tests by varying the number of points in a mesh and number of nodes in the two hidden layers, with the results presented in \TAB{tab1} and \TAB{tab2}. On the one hand, \TAB{tab1} shows the performance of the program by varying the number of points in the mesh, fixing $n_h=16$ for both hidden layers. In order to achieve keV accuracy we need meshes above $1000$ points.
On the other hand,  \TAB{tab2} fixes $M=1000$ and varies $n_{h1}$ and $n_{h2}$, which represent the number of nodes in the first and second hidden layers, respectively. From the table we see that, even though there is a slight fluctuation in each training, $n_h=8$ is enough to obtain the desired precision, so increasing the number of nodes beyond $16$ would not yield higher predictive power.

\begin{table}[htbp]
	\centering
	\caption{Influence of the number of nodes $(n_{h1},n_{h2})$ in different hidden layers on the deuteron results, taking $M=1000$ points in the mesh.}
	\begin{tabular}{cccccc}
		\hline\hline
		$(n_{h1},n_{h2})$& $E_b$ [MeV] & $\langle T\rangle$ [MeV] & $\langle V \rangle$ [MeV] & ${\cal P}_S$ [\%] & ${\cal P}_D$ [\%]\\
		\hline
		$(4,4)$& $2.2041$ & $12.3302$ & $-14.5345$ & $95.71$ & $4.29$\\
		$(8,8)$& $2.2242$ & $11.7190$ & $-13.9423$ & $95.91$ & $4.09$\\
	   $(16,16)$& $2.2241$ & $11.7242$ & $-13.9479$ & $95.90$ & $4.10$\\
		\hline
		\hline
	\end{tabular}%
	\label{tab2}
\end{table}%

\subsection{Three-body system}

\subsubsection{Three-body harmonic oscillator with equal masses}

For the extension to three body systems, as we did for the two body, let's start considering a simple isotropic three body harmonic oscillator with a potential is give by
\begin{align}\label{eq:pot3BHO}
 V = \frac{1}{2}k\left(\rho_1^2+\rho_2^2+\rho_3^2\right)
\end{align}

\begin{figure}[t]
 \includegraphics[width=.5\textwidth]{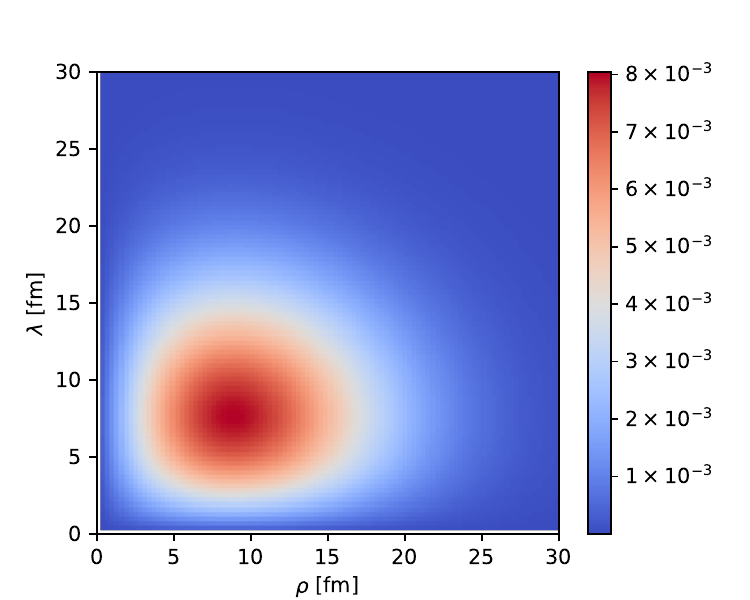}
 \caption{\label{fig:3BHOwf} Predicted reduced radial wave function $u(\rho,\lambda)$ [in fm$^{-1}$] for the three-body isotropic harmonic oscillator.}
\end{figure}

For equal masses this potential is separable,

\begin{align}
 V  = \frac{3}{4}k\rho^2 + k\lambda^2,
\end{align}
and the solution factorizes as $\Psi= \psi_\rho(\rho)\psi_\lambda(\lambda)$. Then, the Schr\"odinger equation of \Eq{eqc12} can be written as,

\begin{equation}
 \begin{aligned}
	\left[-\frac{\hbar^2}{2\mu_{\rho}}\nabla_\rho^2 +\frac{3}{4}k\rho^2\right]\psi_\rho&=E_\rho\psi_\rho,\\
	\left[- \frac{\hbar^2}{2\mu_{\lambda}}\nabla_\lambda^2  + k\lambda^2\right]\psi_\lambda&=E_\lambda\psi_\lambda,
 \end{aligned}
\end{equation}
so the $\psi_{\rho(\lambda)}$ is given by the same solution as in \Eq{eq:2bHOsol} and the energy is,
\begin{align}
 E=E_\rho+E_\lambda=\hbar\omega_\rho\left(n_\rho+\frac{3}{2}\right)+\hbar\omega_\lambda\left(n_\lambda+\frac{3}{2}\right)
\end{align}
with $\omega^2_\rho=\omega^2_\lambda=\frac{3k}{2m}\equiv\omega^2$.

In order to test the three-body code, we will solve the ground state of the potential of \Eq{eq:pot3BHO} for $L=l_\rho=l_\lambda=0$ for three particles with masses $m=1$ GeV and $\hbar\omega=1$ MeV, which has an energy of $E=3$ MeV and where the wave function is given by

\begin{align}\label{eq:analytic3bHO}
 \Psi(\rho,\lambda) &= \left(\frac{2\sqrt{\beta_\rho\beta_\lambda}}{\pi}\right)^{3/2} e^{-\beta_\rho \rho^2-\beta_\lambda\lambda^2}
\end{align}
with $\beta_\rho=250$ MeV$^2$ and $\beta_\lambda=333.3$ MeV$^2$.

\begin{figure}[t]
 \includegraphics[width=.45\textwidth]{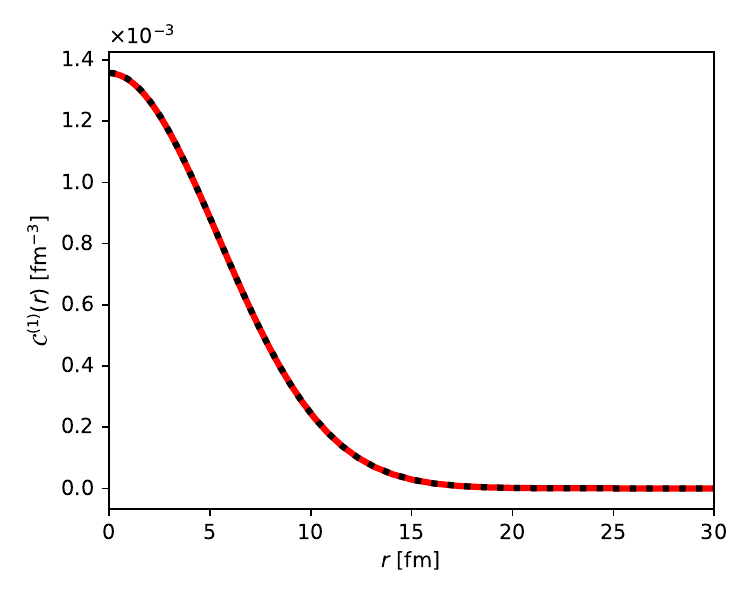}
 \includegraphics[width=.45\textwidth]{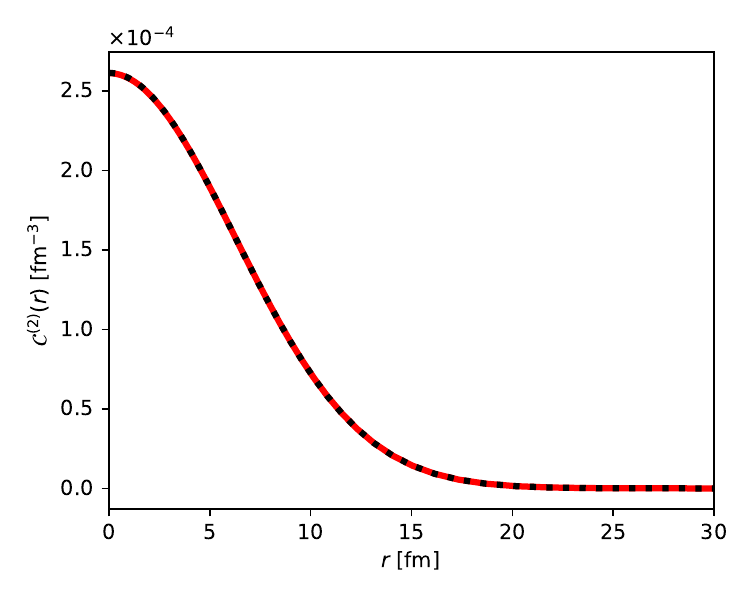}
 \caption{\label{fig:3BHOwf-proj} One-body (upper pannel) and two-body (lower panel) correlation function (defined in Eq.~\eqref{eq:denfun}) from the DNN calculation (red solid line), compared with the exact wave function of Eq.~\eqref{eq:analytic3bHO} (black dotted line) for the three-body isotropic harmonic oscillator.}
\end{figure}

Following the procedure described in Sec.~\ref{sec:threebody}, we build a DNN with two-hidden layers of $8$ nodes each. The $(\rho,\lambda)$ coordinates are discretized using a $M=(100,100)$ mesh between $0.01$ and $30$ fm. As in the two-body case, the softplus function is used as the activation function of the neural network and Adam algorithm is used to optimize the parameters.

The obtained wave function is shown in Fig.~\ref{fig:3BHOwf}. The output of the DNN predicts a wave function which deviates less than $1\%$ of the analytical wave function, predicting an energy of $2.998$ MeV, an average kinetic energy of $\langle T\rangle=1.499$ MeV and an average potential energy $\langle V\rangle=1.499$ MeV.

In Figs.~\ref{fig:3BHOwf-proj} we show the one- and two-body correlation functions, defined as
\begin{equation}\label{eq:denfun}
 \begin{aligned}
 {\cal C}^{(1)}(r) &= \langle \Psi_{JM}|\delta({\bf r}-{\bf r}_{G,i})|\Psi_{JM}\rangle,\\
 {\cal C}^{(2)}(r) &= \langle \Psi_{JM}|\delta({\bf r}-{\boldsymbol \rho}_i)|\Psi_{JM}\rangle,
 \end{aligned}
\end{equation}
where ${\bf r}_{G,i}=\frac{2}{3}{\boldsymbol\lambda}_i$ represents the vector from the CM of the system to the $i$-th particle. As $\Psi_{JM}$ is normalized, ${\cal C}^{(1,2)}(r)$ are normalized as $\int {\cal C}^{(j)}(r)d{\bf r}=1$, for $j=1,2$.
As it can be seen, the neural network solution is in excellent agreement with the analytical solution.

\subsubsection{Study of the Triton}

Finally, in this section we evaluate the validity of the DNN three-body model for coupled channels and more complex potentials involving tensor and spin-orbit terms, by exploring the three-nucleon system. We will consider a $J=T=\tfrac{1}{2}$ state, that's it, the triton case, with the same $\chi$EFT potential employed in the deuteron case of Sec.~\ref{sec:deuteron}. As our aim is to analyze the capability of the neural network for a general potential with coupled channels and not a full accurate calculation of the triton, we restrict ourselves to a calculation including the 3 partial waves listed in \TAB{tabnch}.

\begin{table}[b]
		\captionsetup{justification=raggedright, singlelinecheck=false}
	\caption{Quantum numbers of the partial waves included in the three-nucleon calculation. $\{t_{ij},s_{ij},l_{ij}\}$ are the isospin, spin and orbital momentum of the $(ij)$ pair, respectively, $l_k$ the orbital momentum of the $(ij)-k$ pair and $S$ and $L$ the total spin and orbital momentum of the three nucleons.}
	\begin{tabular}{ccccccc}
		\hline \hline
		nch & $t_{ij}$& $s_{ij}$& $l_{ij}$& $l_k$& $S$&$L$\\
		\hline
		1&1&0&0&0& $\tfrac{1}{2}$ &0\\
		2&0&1&0&0& $\tfrac{1}{2}$ &0\\
		3&0&1&2&0& $\tfrac{3}{2}$ &2\\
		\hline \hline
	\end{tabular}
	\label{tabnch}
\end{table}%

In this case we generate a $(\rho,\lambda)$ mesh in the range $(0,15)$ fm, with $80$ points each. As for the harmonic oscillator,
we use a deep neural network consisting of two $8$-node hidden layers.
For this system we have used $\tanh(x)$ as an activation function in the neural network. Contrary to the softplus function, the $\tanh(x)$ maps input values in the range $[-1,1]$ so the output can be negative, which is convenient for predicting wave functions with nodes or relative phases. This activation function is a popular choice in machine learning techniques as it introduces non-linearities into the network, which is relevant for deep learning models, while it is symmetric around zero, allowing to a better balancing of the weights and making the training faster and more stable.

Additionally, we employ the Adamax optimizer~\cite{Kingma:2014vow} instead of Adam. While Adam scales the learning rate for each parameter proportionally to the inverse of the square root of an exponentially decaying average of past squared gradients, Adamax varies it by using an exponentially decaying average based on the infinity norm of past gradients (in other words, it replaces the $L_2$ norm with the $L_\infty$ norm). In practice, this optimizer can provide greater stability than Adam for specific tasks.

\begin{figure}[t]
 \includegraphics[width=.48\textwidth]{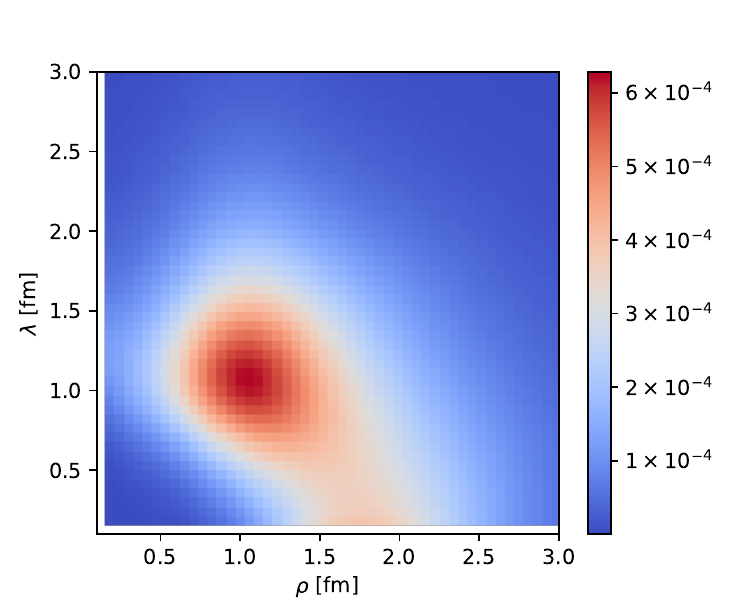}
 \includegraphics[width=.48\textwidth]{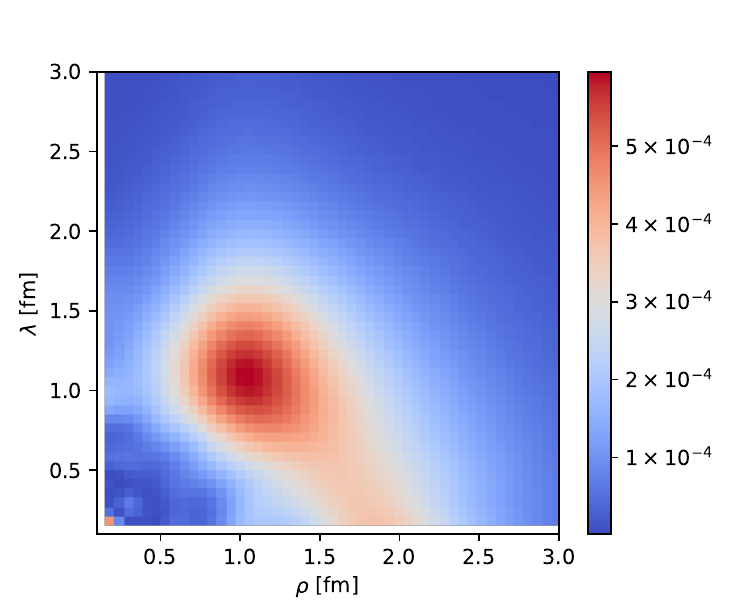}
 \caption{\label{fig:Tritwf} Predicted probability density $\varrho(\rho,\lambda)$ [in fm$^{-6}$] for the triton in the DNN calculation (upper panel) and the GEM calculation (lower panel).}
\end{figure}

The triton binding energy obtained is ${\cal B}_{\rm t}=9.83$ MeV, which is slightly larger than the experimental value ${\cal B}^{(\rm exp)}_{\rm t}=8.481795$ MeV~\cite{CPC:10.1088/1674-1137/41/3/030003} and the binding energy given in Ref.~\cite{Saha:2022oep}, of ${\cal B}^{(\chi{\rm EFT})}_{\rm t}=8.09$ MeV at N$^3$LO. The reason is that the triton calculation in Ref.~\cite{Saha:2022oep} is done with 34 channels using a $\chi$EFT $NN$ potential that includes ${\bf L}^2$ and $({\bf L}\cdot{\bf S})^2$ terms, which are not included in this work and reduce the binding energy. Hence, in order to compare similar scenarios, we have solved the three-nucleon bound state with the three partial waves in Tab.~\ref{tabnch} with GEM using the same channels and a potential that includes the central, tensor and spin-orbit terms. In this case, we obtain ${\cal B}_{\rm t}^{(\rm GEM)}=9.77$ MeV, only $0.6\%$ away from the DNN result.
The average kinetic and potential energies obtained are $\langle T\rangle = 44.9$ MeV and $\langle V \rangle =-54.7$ MeV, which are in good agreement with the GEM results, $\langle T\rangle_{\rm GEM}=44.4$ MeV and $\langle V\rangle_{\rm GEM}=-54.2$ MeV.
In addition, we have calculated the $D$ channel probability, obtaining ${\cal P}_D=6.67$ \% for the DNN, whereas the GEM result gives $6.70$ \%.

\begin{figure}[t]
 \includegraphics[width=.45\textwidth]{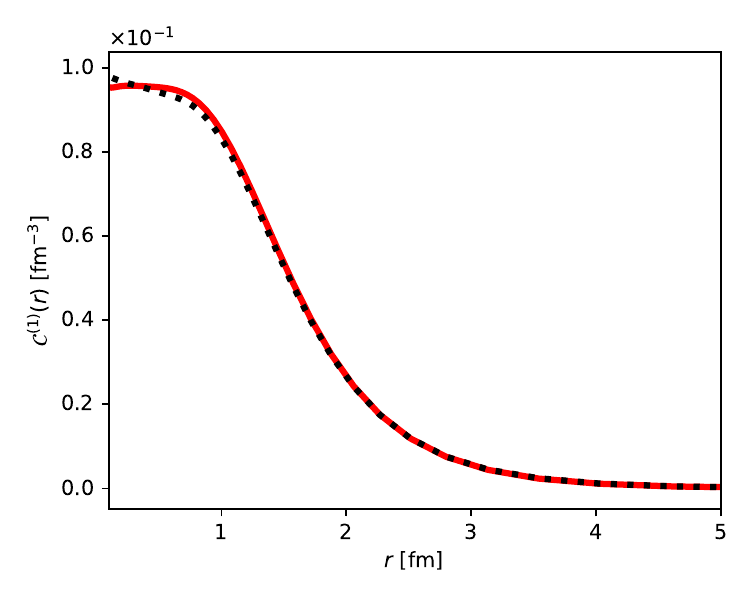}
 \includegraphics[width=.45\textwidth]{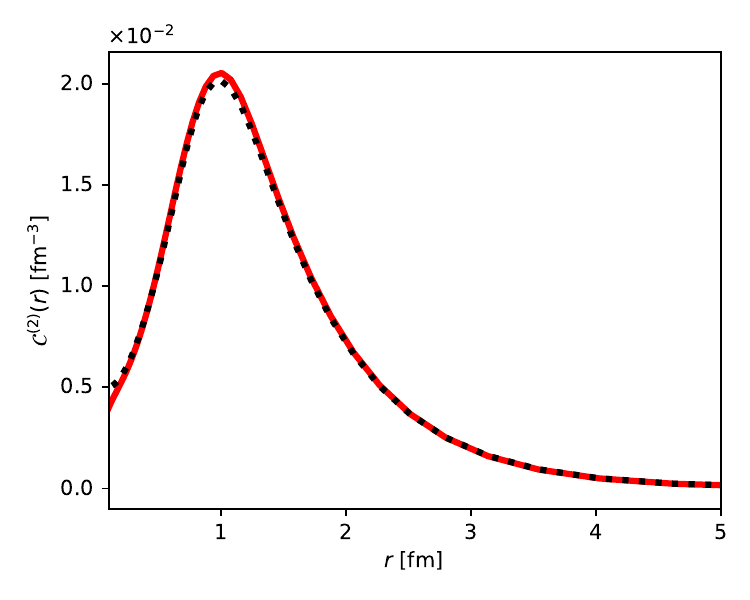}
 \caption{\label{fig:Trit-proj} One-body (upper pannel) and two-body (lower panel) correlation functions (defined in Eq.~\eqref{eq:denfun}) from the DNN calculation (red line), compared with the result obtained from the Gaussian Expansion Method (black dotted line) for the triton.}
\end{figure}

In Fig.~\ref{fig:Tritwf} we show the probability density of the triton obtained with the DNN code, defined as,

\begin{align}
\varrho(\rho,\lambda) &= \int d\Omega_\rho\,d\Omega_\lambda\,|\Psi({\boldsymbol\rho,\boldsymbol\lambda)}|^2,
\end{align}
compared to the GEM result using a 196-Gaussian base and $\{\rho_{\rm min},\rho_{\rm max},\lambda_{\rm min},\lambda_{\rm max}\} = \{0.2,12,0.2,9\}$ fm. We find a good agreement, except for the low $\rho$ and $\lambda$ values, where the GEM result shows an oscillation we associate with the Gaussian basis with very small ranges.

Figs.~\ref{fig:Trit-proj} show the one- and two-body correlation functions compared to the GEM calculation. We see that both curves agree, with only a small deviation at the very short-range.
Finally, we have calculated the mass root-mean-square radius which can be derived from the one-body correlation function as

\begin{align}
 \bar r_m^2 = \int r^2\,{\cal C}^{(1)}(r) d{\bf r}.
\end{align}
For the triton, we obtain $\bar r_m=1.65$ fm for the DNN calculation and $1.73$ fm for GEM.

As we have found, the solution of the three nucleon system solved with the multipole expansion and the deep neural network provides a solution compatible with other well-known methods such as the Gaussian Expansion method, with no need to give any ansatz on the trial wave function which, in some cases, provides a better-behaved solution.


 \section{Summary}
 \label{IV}

 In this paper we have constructed a deep neural network (DNN) method to solve the general two and three body systems. This approach works by taking the coordinates as the input layer of DNN with no need to parameterize the initial wave function by means of trial functions, and it is applicable to any type of potential and number of channels. In particular, we use a novel multipole expansion method to solve the three-body systems, which involves more complexity than the two body states.
We have tested two different potential models: the harmonic oscillator potential and a realistic $NN$ potential based on the chiral effective field theory ($\chi$EFT) approach~\cite{Saha:2022oep}, to verify the validity of the method and employed it with two-body and three-body systems, respectively.

Firstly, we evaluated the DNN result by comparing the wave function and the exact solution in the two-body isotropic harmonic oscillator. Numerical calculations show that the deviation is less than 2\%. Subsequently, we applied the method to deuteron studies and the results were in agreement with the experimental and theoretical values. Furthermore, we compared the wave functions obtained by DNN and Gaussian Expansion Method (GEM) and they showed consistent line behavior. By varying the number of points in a mesh and the number of nodes in the hidden layers, we also analyzed the convergence of the deep neural network results. 

Secondly, when extending the analysis to the three-body system, we initially studied harmonic oscillator and showed that the DNN wave function deviates from the analytical result within 1\%. We, then, compared the properties of triton obtained from DNN predictions with GEM results, finding a good agreement between both methods.

Finally, we believe that this method represents a promising initial step towards applying deep neural networks to other many-body systems, since the study can be generalised to systems comprising more than three particles. This paves the way for future studies of four-body systems. However, extending this work to cover such systems is beyond the scope of this study and will have to be the subject of future research.


\begin{acknowledgments}
 	\noindent
P.G.O. wants to thank M.~Kamimura and D.R.~Entem for fruitful discussions.
Work partially financed by
Project No. PID2022-141910NB-I00 funded by Spanish MICIU/AEI/10.13039/501100011033;
Junta de Castilla y León program EDU/841/2024 under grant No. SA091P24 under the title: ``Las nuevas tecnologías: computación cuántica y aprendizaje automático para estudiar las interacciones fundamentales y sus aplicaciones a la física médica".
H.S. is supported by the National Natural Science Foundation of China (Grant No.12075043). 
X.L. is supported
by the National Natural Science Foundation of China under Grant No.12205002.
\end{acknowledgments}

\appendix
\section{Multipole expansion of the potential for the three-body system}\label{sec:multipoleNN}

The multipole expansion of the two-body $NN$ potential has been studied in, e.g., Refs.~\cite{Horie1961,Umeya:2016oqt}, but its application in the context of the three-body system is new, up to our knowledge. In this appendix we will give  a brief description.
Let's assume we have a general potential $V({\bf r})$ that can be written as a product of spin and spatial parts,

\begin{align}
    V_k({\bf r}) &= V(r)\,{\cal O}_S^{(k)}\cdot{\cal O}_L^{(k)}
\end{align}
where ${\cal O}_{S(L)}^{(k)}$ is an irreducible tensor operator of order $k$ in the orbital (spin) space of two interacting particles. The matrix elements can be factorized in spatial and spin parts as,
\begin{align}
    \langle L'S';J'M'|&{\cal O}_S^{(k)}\cdot{\cal O}_L^{(k)}|LS;JM\rangle = \delta_{JJ'}\delta_{MM'}(-1)^{L+S'+J}\nonumber\\&
    \Sixj{L}{S}{J}{S'}{L'}{k}\langle S'||{\cal O}_S^{(k)}||S\rangle \langle L'||{\cal O}_L^{(k)}||L\rangle
\end{align}
 which depends on the Wigner's 6-j symbol and
 where $\langle L'||{\cal O}_L^{(k)}||L\rangle$ and $\langle S'||{\cal O}_S^{(k)}||S\rangle$ are the reduced matrix elements. These can be calculated through the Wigner-Eckart theorem (for the spatial part is similar),

 \begin{align}
 \langle S'M_S'|{\cal O}_S^{(k,m_k)}|SM\rangle &= A\,\langle S'||{\cal O}_S^{(k)}||S\rangle  \end{align}
 with,
 \begin{align}
     A &= (-1)^{S'-M_S'}\Threej{S'}{k}{S}{-M_S'}{m_k}{M_S}
 \end{align}
 which depends on the Wigner's 3-j symbol.

 Focusing on the spatial part, $V_k(r){\cal O}^{(k)}_L$, the central, tensor and spin-orbit $NN$ interaction can be generally summarized as,

 \begin{align}\label{eq:genpoten}
     V({\bf r}){\cal O}_L^{(k)} &\equiv V_k(r)\,{\cal C}_k(\hat r)
 \end{align}
 where $V_k(r)$ is a radial function and ${\cal C}_k$ is a rank-$k$ spherical tensor operator defined as,
 \begin{align}
     {\cal C}_{k,m_k}(\hat r) &= \frac{\sqrt{4\pi}}{\hat k} Y_{k,m_k}(\hat r)
 \end{align}
 begin $\hat{k}\equiv\sqrt{2k+1}$ and $Y_{k,m_k}(\hat r)$ the spherical harmonic function.
 For example, the central, tensor and spin-orbit $NN$ interactions can be written as a product of spin(isospin) and spatial operators, as~\cite{Umeya:2016oqt},

\begin{align}
     V_C({\bf r}) &= V_0(r)\left\{\begin{array}{c}1\\\sigma_1\cdot\sigma_2\end{array}\right\}{\cal C}_0(\hat r),\\
     V_T({\bf r})S_{12} &= \sqrt{\frac{2}{3}}\left([\sigma_1\otimes\sigma_2]^{(2)}\cdot V_2(r){\cal C}_2(\hat r)\right),\\
     V_{LS}({\bf r})({\bf S}\cdot{\bf L}) &= -i\sqrt{2}\left( {\bf S}\cdot\left[r\,V_1(r){\cal C}_1(\hat r)\times {\bf p}\right]_1\right).\label{eq:LSpot}
 \end{align}


First we will focus on the central and tensor part. If we assume the $V_k({\bf r})$ potential of Eq.~\eqref{eq:genpoten} has a well-defined Fourier transform, we can write~\cite{Weniger1985},

\begin{align}
 \tilde{V}_k({\bf p}) &= (2\pi)^{-3/2}\,\int e^{-i\,{\bf p}\cdot{\bf r}}V_k({\bf r})d^3{\bf r},\\
 V_k({\bf r}) &= (2\pi)^{-3/2}\,\int e^{i\,{\bf p}\cdot{\bf r}}\tilde{V}_k({\bf p})d^3{\bf p}. \label{eq:four}
\end{align}

If we now want to calculate $V({\bf r})$ in terms of two variables ${\bf r}_1$ and ${\bf r}_2$, with ${\bf r}={\bf r}_1+{\bf r}_2$ we have,

\begin{align}\label{eq:FT}
 V_k({\bf r}) &= (2\pi)^{-3/2}\,\int e^{i\,{\bf p}\cdot(\,{\bf r}_1+{\bf r}_2)}\tilde{V}_k({\bf p})d^3{\bf p}.
\end{align}

As $V_k({\bf r})$ depends on the spherical irreducible tensor ${\cal C}_k(\hat r)$, in order to calculate the Fourier transform we use the well-known Rayleigh expansion of a plane wave in terms of spherical Bessel functions and spherical harmonics,

\begin{align}
 e^{\pm\,i\,{\bf p}\cdot{\bf r}} &= 4\pi\,\sum_{\ell,m}\,(\pm i)^\ell\,j_\ell(pr) Y_{\ell m}(\hat r)Y^*_{\ell m}(\hat p).
\end{align}

Then, using this expansion in Eq.~\eqref{eq:FT}, we end up with ,

\begin{align}
 V_k({\bf r}) &= \sqrt{32\pi}\sum_{l_1,m_1}\sum_{l_2,m_2}\,i^{l_1+l_2}\,\,Y_{l_1m_1}(\hat r_1)Y_{l_2m_2}(\hat r_2)\times\nonumber\\&
 \times\int \,j_{l_1}(r_1 p)j_{l_2}(r_2p) \tilde{V}_k({\bf p})Y^*_{l_1m_1}(\hat p)Y^*_{l_2m_2}(\hat p)d^3{\bf p}.
\end{align}

Now, using the inverse Fourier transform for  $\tilde V({\bf p})$ and expanding again the exponential, we end up with

\begin{align}
 V_k({\bf r}) &= 8(-1)^{k}\sum_{l_1,m_1}\sum_{l_2,m_2}\,i^{l_1+l_2+k}\,\,Y_{l_1m_1}(\hat r_1)Y_{l_2m_2}(\hat r_2)\nonumber\times\\&
 \times\langle k m_k|l_1m_1|l_2m_2\rangle\int {\cal I}_{(l_1,l_2,k)}(r_1,r_2,r')\,V_k(r')\,r^{\prime\,2}dr',
\end{align}
where we have defined,
\begin{align}
 \langle \ell m|l_1m_1|l_2m_2\rangle &= \int d\Omega\, Y_{\ell m} Y^*_{l_1m_1} Y^*_{l_2m_2}
\end{align}
and

\begin{align}
 {\cal I}_{(l_1,l_2,k)}(r_1,r_2,r') &= \int_0^\infty j_{l_1}(r_1p)j_{l_2}(r_2p)j_k(pr') p^2 dp.
\end{align}

This integral involves three spherical Bessel functions and, fortunately, can be integrated~\cite{Mehrem:1990eg,Mehrem:2010qk}. So we can finally express the potential as,

\begin{align}\label{eq:multipoleexpansion}
 V_k(|{\bf r}_1+{\bf r}_2|) &= \sum_{l_1,l_2}\,V_{l_1,l_2;k}(r_1,r_2)\,[Y_{l_1}(\hat r_1)\otimes Y_{l_2}(\hat r_2)]_k
\end{align}
where

\begin{align}
 V_{l_1,l_2;k}(r_1,r_2) &= \sum_{{\cal L},\ell} r_1^{k-{\cal{L}}}r_2^{\cal{L}}\,{\cal A}_{({\cal L},\ell;l_1,\,l_2,\,k)}\cdot v_{(k,\ell)}
\end{align}
with $ r= \sqrt{r_1^2+r_2^2+2\,r_1\,r_2\,z}$ and where we have defined,

\begin{align}\label{eq:calA}
&{\cal A}_{({\cal L},\ell;l_1,l_2,k)}
=\,\sqrt{4\pi}(-1)^{l_1+l_2+\ell}\, \hat{k}\hat{l_1}\hat{l_2}\hat{\ell}^2
\,\Comb{2k}{2{\cal L}}^{1/2}\times\nonumber\\&
\times\Threej{l_1}{k - {\cal L}}{\ell}{0}{0}{0}
\Threej{l_2}{{\cal L}}{\ell}{0}{0}{0}\,
\Sixj{l_1}{l_2}{k}{{\cal L}}{k-{\cal L}}{\ell}.
\end{align}
and
\begin{align}
v_{(k,\ell)} \equiv v_{(k,\ell)}(r_1,r_2) =\frac{1}{2}\int_{-1}^1 \frac{V_k(r)}{r^k}\,P_\ell(z)dz
\end{align}
with $P_\ell(z)$ the Legendre polynomial of order $\ell$.

The spin-orbit part is a bit trickier, but it can be done in a similar way. Following Ref.~\cite{Horie1961} we will split the spatial part of $V_{LS}({\bf r})$ in Eq.~\eqref{eq:LSpot} as,

\begin{align}
V(r)&[{\boldsymbol \rho}_c \times {\bf q}_c] =
 {\cal V}^{(1)} - {\cal V}^{(2)} \frac{\alpha_{c1}\rho}{\beta_{c1}\lambda}-{\cal V}^{(3)} \frac{\beta_{c1}\lambda}{\alpha_{c1}\rho}+\nonumber\\&
 +{\cal V}^{(4)}\,\left\{\frac{\beta_{c1}}{\alpha_{c1}}\lambda\frac{\partial}{\partial{\rho}}-\frac{\alpha_{c1}}{\beta_{c1}}\rho\frac{\partial}{\partial{\lambda}}\right\}.
\end{align}
The ${\cal V}^{(i)}$ terms are given by

\begin{equation}
\begin{aligned}
{\cal V}^{(1)} =& \sum_\ell(-1)^\ell \frac{\hat{\ell}^2}{\sqrt{3}}v_{(\ell)}\left[\hat{\ell}\left\{{\cal D}_{(\ell,\ell)}^{(1)} -{\cal D}_{(\ell,\ell)}^{(2)}\right\}+\right.\nonumber\\&+\left.\sum_{\ell'=\ell\pm 1} \hat{\ell'}\left\{
{\cal D}_{(\ell,\ell')}^{(1)} + {\cal D}_{(\ell,\ell')}^{(2)}\right\}\right]
,\\
{\cal V}^{(2)} =& -\sum_{\ell}(-1)^\ell \frac{\hat{\ell}}{\sqrt{3}}\left[ \left\{ (\ell+1)v_{(\ell-1)}+\ell v_{(\ell+1)} \right\}
{\cal D}_{(\ell,\ell)}^{(1)}+\right.\nonumber\\&\left.+\sum_{\ell'=\ell\pm1}\hat{\ell}\hat{\ell'} v_{(\ell')}
{\cal D}_{(\ell,\ell')}^{(1)} \right] ,\\
{\cal V}^{(3)} =& \sum_{\ell}(-1)^\ell \frac{\hat{\ell}}{\sqrt{3}}\left[ \left\{(\ell+1)v_{(\ell-1)}+\ell v_{(\ell+1)}\right\}
{\cal D}_{(\ell,\ell)}^{(2)}-\right.\nonumber\\&-\left.\sum_{\ell'=\ell\pm1}\hat{\ell}\hat{\ell'}v_{(\ell')}{\cal D}_{(\ell,\ell')}^{(2)} \right],\\
 {\cal V}^{(4)} =& \sum_\ell (-1)^\ell\sqrt{\frac{\ell(\ell+1)(2\ell+1)}{3}}\{v_{(\ell-1)}-v_{(\ell+1)}\}\nonumber\\&[{\cal C}_\ell(\hat r_1)\otimes {\cal C}_\ell(\hat r_2)]_1,\\
\end{aligned}
\end{equation}
where $v_{(\ell)}\equiv v_{(1,\ell)}$ and

\begin{equation}
    \begin{aligned}
 {\cal D}_{(\ell,\ell')}^{(1)} &= [[{\cal C}_\ell(\hat r_1)\otimes {\bf L}_{(1)}]_{\ell'} \otimes {\cal C}_\ell(\hat r_2)]_1,\\
 {\cal D}_{(\ell,\ell')}^{(2)} &= [{\cal C}_\ell(\hat r_1) \otimes [{\cal C}_\ell(\hat r_2)\otimes {\bf L}_{(2)}]_{\ell'}]_1.
    \end{aligned}
\end{equation}

This expression may seem complicated at first sight, but the irreducible matrix elements can be easily calculated with,

\begin{equation}
    \begin{aligned}
 \langle l'|| [{\cal C}_{\ell}\otimes {\bf L}]_{\ell'} ||l\rangle &= \sqrt{2l(l+1)}\hat{l}\langle \ell0,11|\ell'1\rangle \langle l1,\ell'(-1)|l'0\rangle,\\
 \langle l'|| {\cal C}_\ell ||l\rangle &= \hat{l}\,\langle l0,\ell0|l'0\rangle,
    \end{aligned}
\end{equation}
with $\ell'=\{\ell,\ell\pm1\}$ and $\langle l_1m_1,l_2m_2|lm\rangle$ the Clebsch-Gordan coefficient.

\section{Multipole expansion of the three body wave function  }\label{eq:multiplePsi}

The three-body wave function in channel $c$ can also be expanded in the $({\boldsymbol \rho},{\boldsymbol \lambda})$ system, similarly to the multiple expansion of the potential detailed in previous section.
Let's assume we have $\psi^{(c)}({\boldsymbol \rho}_c,{\boldsymbol \lambda}_c)$,

\begin{align}
 \psi^{(c)}({\boldsymbol \rho}_c,{\boldsymbol \lambda}_c)=\psi(\rho_c,\lambda_c) [Y_{l_\rho}(\hat \rho_c) Y_{l_\lambda}(\hat\lambda_c)]_{LM}
\end{align}
and we want to calculate $\psi^{(c)}({\boldsymbol \rho},{\boldsymbol \lambda})$ with ${\boldsymbol \rho}_c = \alpha\,{\boldsymbol \rho}+\beta\,{\boldsymbol \lambda}$, ${\boldsymbol \lambda}_c = \gamma\,{\boldsymbol \rho}+\delta\,{\boldsymbol \lambda}$.
Then, using the Fourier transform in each variable we can write,

\begin{align}
 \psi({\boldsymbol \rho}_c,{\boldsymbol \lambda}_c) = (2\pi)^{-3}\,\int &e^{i\,{\bf k}_\rho\cdot(\alpha\,{\boldsymbol \rho}+\beta\,{\boldsymbol \lambda})+i\,{\bf k}_\lambda\cdot(\gamma\,{\boldsymbol \rho}+\delta\,{\boldsymbol \lambda})}\times\nonumber\\
 &\times\tilde{\psi}({\bf k}_\rho,{\bf k}_\lambda)d^3{\bf k}_\rho d^3{\bf k}_\lambda.
\end{align}

Following the same procedure as for the potential, we use the Rayleigh expansion for each exponential and the inverse Fourier transform, obtaining the compact formula,

\begin{align}
 \psi({\boldsymbol \rho}_c,{\boldsymbol \lambda}_c) &=
  \sum_{l_i} \Phi(l_i) \overline{\psi}(l_x,l_y;\rho,\lambda)\left[Y_{l_a}(\hat \rho)Y_{l_b}(\hat \lambda)\right]_{LM}
\end{align}
where $l_i\equiv\{l_1,l_2,l_3,l_4,l_a,l_b,l_x,l_y,{\cal L}_x,{\cal L}_y\}$. Here, we have also defined,

\begin{align}\label{eq:PhiME}
 \Phi(l_i) &= (\alpha\rho)^{l_\rho-{\cal L}_x}(\beta\lambda)^{{\cal L}_x}
 (\gamma\rho)^{l_\lambda-{\cal L}_y}(\delta\lambda)^{{\cal L}_y} B_{l_1,l_2}^{l_a}B_{l_3,l_4}^{l_b}\,
 \times\nonumber\\
 &\times
 \hat{l_\rho}\hat{l_\lambda}\hat{l_a}\hat{l_b}\Ninej{l_1}{l_3}{l_\rho}{l_2}{l_4}{l_\lambda}{l_a}{l_b}{L}{\cal A}_{({\cal L}_x,l_x;l_1,\,l_3,\,l_\rho)}{\cal A}_{({\cal L}_y,l_y;l_2,\,l_4,\,l_\lambda)},
\end{align}
with
\begin{align}
 B_{l_1,l_2}^\ell &= (-1)^\ell \frac{\hat{l_1}\hat{l_2}}{\sqrt{4\pi}}\Threej{l_1}{l_2}{\ell}{0}{0}{0}
\end{align}
and with ${\cal A}$ defined in Eq.~\eqref{eq:calA}. Finally, we define

\begin{align}\label{eq:PsiME}
\overline{\psi}(l_x,l_y;\rho,\lambda) &= \frac{1}{4}\int_{-1}^1 dz\int_{-1}^1 dz' \,\frac{\psi(x,y)}{x^{l_\rho}y^{l_\lambda}}P_{l_x}(z)P_{l_y}(z')
\end{align}
with $P_\ell(z)$ the Legendre polynomial of order $\ell$, and
\begin{align*}
 x &= \sqrt{(\alpha\rho)^2+(\beta\lambda)^2+2(\alpha\rho)(\beta\lambda)z},\\
 y &= \sqrt{(\gamma\rho)^2+(\delta\lambda)^2+2(\gamma\rho)(\delta\lambda)z'}.\\
\end{align*}

\bibliography{ref}
\end{document}